\pdfoutput=1

\documentclass[12pt,a4paper]{article}

\usepackage{ifthen} 
\newboolean{pdflatex}
\setboolean{pdflatex}{true} 

\newboolean{articletitles}
\setboolean{articletitles}{true} 

\newboolean{uprightparticles}
\setboolean{uprightparticles}{false} 

\newboolean{inbibliography}
\setboolean{inbibliography}{false} 

\newboolean{prlstyle}
\setboolean{prlstyle}{false}

\usepackage{microtype}
\usepackage{lineno}  
\usepackage{xspace} 
\usepackage{caption} 

\usepackage{graphicx}  
\usepackage{color}
\usepackage{colortbl}
\graphicspath{{./figs/}{}} 

\usepackage{amsmath} 
\usepackage{amssymb}
\usepackage{amsfonts}
\usepackage{upgreek} 

\newcommand*\patchAmsMathEnvironmentForLineno[1]{%
\expandafter\let\csname old#1\expandafter\endcsname\csname #1\endcsname
\expandafter\let\csname oldend#1\expandafter\endcsname\csname
end#1\endcsname
 \renewenvironment{#1}%
   {\linenomath\csname old#1\endcsname}%
   {\csname oldend#1\endcsname\endlinenomath}%
}
\newcommand*\patchBothAmsMathEnvironmentsForLineno[1]{%
  \patchAmsMathEnvironmentForLineno{#1}%
  \patchAmsMathEnvironmentForLineno{#1*}%
}
\AtBeginDocument{%
\patchBothAmsMathEnvironmentsForLineno{equation}%
\patchBothAmsMathEnvironmentsForLineno{align}%
\patchBothAmsMathEnvironmentsForLineno{flalign}%
\patchBothAmsMathEnvironmentsForLineno{alignat}%
\patchBothAmsMathEnvironmentsForLineno{gather}%
\patchBothAmsMathEnvironmentsForLineno{multline}%
\patchBothAmsMathEnvironmentsForLineno{eqnarray}%
}

\usepackage{hyperref}    
\usepackage[all]{hypcap} 

\usepackage{xspace} 
\usepackage{upgreek}

\def\lhcb {\mbox{LHCb}\xspace}

\def\MagUp {\mbox{\em Mag\kern -0.05em Up}\xspace}

\ifthenelse{\boolean{uprightparticles}}%
{

 \def\Ppi         {\ensuremath{\uppi}\xspace}

 \def\PDelta      {\ensuremath{\Delta}\xspace}                 
 \def\PXi      {\ensuremath{\Xi}\xspace}                 
 \def\PLambda      {\ensuremath{\Lambda}\xspace}                 
 \def\PSigma      {\ensuremath{\Sigma}\xspace}                 
 \def\POmega      {\ensuremath{\Omega}\xspace}                 
 \def\PUpsilon      {\ensuremath{\Upsilon}\xspace}                 
 
 \def\PB      {\ensuremath{\mathrm{B}}\xspace}                 
                  
 \def\PD      {\ensuremath{\mathrm{D}}\xspace}

 \def\PK      {\ensuremath{\mathrm{K}}\xspace}

 \def\Pb      {\ensuremath{\mathrm{b}}\xspace}                 
 \def\Pc      {\ensuremath{\mathrm{c}}\xspace}

 \def\Pi      {\ensuremath{\mathrm{i}}\xspace}

 \def\Ps      {\ensuremath{\mathrm{s}}\xspace}

}
{

 \def\Ppi         {\ensuremath{\pi}\xspace}

 \mathchardef\PDelta="7101
 \mathchardef\PXi="7104
 \mathchardef\PLambda="7103
 \mathchardef\PSigma="7106
 \mathchardef\POmega="710A
 \mathchardef\PUpsilon="7107
                  
 \def\PB      {\ensuremath{B}\xspace}                 
                  
 \def\PD      {\ensuremath{D}\xspace}

 \def\PK      {\ensuremath{K}\xspace}

 \def\Pb      {\ensuremath{b}\xspace}                 
 \def\Pc      {\ensuremath{c}\xspace}

 \def\Pi      {\ensuremath{i}\xspace}

 \def\Ps      {\ensuremath{s}\xspace}

}

\makeatletter
\ifcase \@ptsize \relax
  \newcommand{\miniscule}{\@setfontsize\miniscule{4}{5}}
\or
  \newcommand{\miniscule}{\@setfontsize\miniscule{5}{6}}
\or
  \newcommand{\miniscule}{\@setfontsize\miniscule{5}{6}}
\fi
\makeatother

\DeclareRobustCommand{\optbar}[1]{\shortstack{{\miniscule (\rule[.5ex]{1.25em}{.18mm})}
  \\ [-.7ex] $#1$}}

\def\squark    {{\ensuremath{\Ps}}\xspace}

\def\cquark    {{\ensuremath{\Pc}}\xspace}

\def\bquark    {{\ensuremath{\Pb}}\xspace}

\def\pion   {{\ensuremath{\Ppi}}\xspace}
\def\piz    {{\ensuremath{\pion^0}}\xspace}

\def\pip    {{\ensuremath{\pion^+}}\xspace}
\def\pim    {{\ensuremath{\pion^-}}\xspace}

\def\kaon    {{\ensuremath{\PK}}\xspace}
  \def\Kbar    {{\kern 0.2em\overline{\kern -0.2em \PK}{}}\xspace}

\def\KorKbar    {\kern 0.18em\optbar{\kern -0.18em K}{}\xspace}
\def\Kz      {{\ensuremath{\kaon^0}}\xspace}
\def\Kzb     {{\ensuremath{\Kbar{}^0}}\xspace}
\def\Kp      {{\ensuremath{\kaon^+}}\xspace}

\def\KS      {{\ensuremath{\kaon^0_{\mathrm{ \scriptscriptstyle S}}}}\xspace}

\def\Kstarz  {{\ensuremath{\kaon^{*0}}}\xspace}

\def\Kstarp  {{\ensuremath{\kaon^{*+}}}\xspace}

  \def\Dbar    {{\kern 0.2em\overline{\kern -0.2em \PD}{}}\xspace}

\def\DorDbar    {\kern 0.18em\optbar{\kern -0.18em D}{}\xspace}

\def\Dzb     {{\ensuremath{\Dbar{}^0}}\xspace}

\def\Dstarzb {{\ensuremath{\Dbar{}^{*0}}}\xspace}

\def\B       {{\ensuremath{\PB}}\xspace}
\def\Bbar    {{\ensuremath{\kern 0.18em\overline{\kern -0.18em \PB}{}}}\xspace}

\def\BorBbar    {\kern 0.18em\optbar{\kern -0.18em B}{}\xspace}
\def\Bz      {{\ensuremath{\B^0}}\xspace}
\def\Bzb     {{\ensuremath{\Bbar{}^0}}\xspace}
\def\Bu      {{\ensuremath{\B^+}}\xspace}

\def\Bp      {{\ensuremath{\Bu}}\xspace}

\def\Bd      {{\ensuremath{\B^0}}\xspace}
\def\Bs      {{\ensuremath{\B^0_\squark}}\xspace}

\def\Bdb     {{\ensuremath{\Bbar{}^0}}\xspace}

  \def\Y#1S{\ensuremath{\PUpsilon{(#1S)}}\xspace}

\def\Lbar        {{\ensuremath{\kern 0.1em\overline{\kern -0.1em\PLambda}}}\xspace}
\def\LorLbar    {\kern 0.18em\optbar{\kern -0.18em \PLambda}{}\xspace}

\def\BF         {{\ensuremath{\mathcal{B}}}\xspace}

\def\BR         {\BF}
\newcommand{\decay}[2]{\ensuremath{#1\!\to #2}\xspace}         

\def\to                 {\ensuremath{\rightarrow}\xspace}

\def\CP                {{\ensuremath{C\!P}}\xspace}

\def\AT#1     {\ensuremath{A_{\mathrm{T}}^{#1}}\xspace}           

\def\C#1      {\ensuremath{\mathcal{C}_{#1}}\xspace}                       
\def\Cp#1     {\ensuremath{\mathcal{C}_{#1}^{'}}\xspace}                    
\def\Ceff#1   {\ensuremath{\mathcal{C}_{#1}^{\mathrm{(eff)}}}\xspace}        
\def\Cpeff#1  {\ensuremath{\mathcal{C}_{#1}^{'\mathrm{(eff)}}}\xspace}       
\def\Ope#1    {\ensuremath{\mathcal{O}_{#1}}\xspace}                       
\def\Opep#1   {\ensuremath{\mathcal{O}_{#1}^{'}}\xspace}                    

\newcommand{\tev}{\ifthenelse{\boolean{inbibliography}}{\ensuremath{~T\kern -0.05em eV}\xspace}{\ensuremath{\mathrm{\,Te\kern -0.1em V}}}\xspace}
\newcommand{\gev}{\ensuremath{\mathrm{\,Ge\kern -0.1em V}}\xspace}
\newcommand{\mev}{\ensuremath{\mathrm{\,Me\kern -0.1em V}}\xspace}
\newcommand{\kev}{\ensuremath{\mathrm{\,ke\kern -0.1em V}}\xspace}
\newcommand{\ev}{\ensuremath{\mathrm{\,e\kern -0.1em V}}\xspace}
\newcommand{\gevc}{\ensuremath{{\mathrm{\,Ge\kern -0.1em V\!/}c}}\xspace}
\newcommand{\mevc}{\ensuremath{{\mathrm{\,Me\kern -0.1em V\!/}c}}\xspace}
\newcommand{\gevcc}{\ensuremath{{\mathrm{\,Ge\kern -0.1em V\!/}c^2}}\xspace}
\newcommand{\gevgevcccc}{\ensuremath{{\mathrm{\,Ge\kern -0.1em V^2\!/}c^4}}\xspace}
\newcommand{\mevcc}{\ensuremath{{\mathrm{\,Me\kern -0.1em V\!/}c^2}}\xspace}

\def\invfb   {\ensuremath{\mbox{\,fb}^{-1}}\xspace}

\def\ps   {\ensuremath{{\mathrm{ \,ps}}}\xspace}

\newcommand{\stat}{\ensuremath{\mathrm{\,(stat)}}\xspace}
\newcommand{\syst}{\ensuremath{\mathrm{\,(syst)}}\xspace}

\newcommand{\chisq}{\ensuremath{\chi^2}\xspace}

\def\gsim{{~\raise.15em\hbox{$>$}\kern-.85em
          \lower.35em\hbox{$\sim$}~}\xspace}
\def\lsim{{~\raise.15em\hbox{$<$}\kern-.85em
          \lower.35em\hbox{$\sim$}~}\xspace}

\def\ptot       {\mbox{$p$}\xspace}
\def\pt         {\mbox{$p_{\mathrm{ T}}$}\xspace}

\def\degrees{\ensuremath{^{\circ}}\xspace}

\def\rad{\ensuremath{\mathrm{ \,rad}}\xspace}

\def\evtgen     {\mbox{\textsc{EvtGen}}\xspace}

\def\geant      {\mbox{\textsc{Geant4}}\xspace}

\def\photos     {\mbox{\textsc{Photos}}\xspace}

\def\pythia     {\mbox{\textsc{Pythia}}\xspace}

\def\tell1  {TELL1\xspace}
\def\ukl1   {UKL1\xspace}

\usepackage{cite} 
\usepackage{mciteplus}

\usepackage{longtable} 

\usepackage{booktabs}
\usepackage{multirow}

\def\Dstarzbprime {{\ensuremath{\Dbar{}^{(*)0}}}\xspace}

\def\Bsum         {{\ensuremath{\mathcal{B}_{\mathrm{sum}}}}\xspace}
\def\BorBs         {{\ensuremath{B_{(s)}^0}}\xspace}

\newcommand{\systfsfd}{\ensuremath{\mathrm{\,(frag)}}\xspace}
\newcommand{\systext}{\ensuremath{\mathrm{\,(norm)}}\xspace}

\begin{document}

\renewcommand{\thefootnote}{\fnsymbol{footnote}}
\setcounter{footnote}{1}


\begin{titlepage}
\pagenumbering{roman}

\vspace*{-1.5cm}
\centerline{\large EUROPEAN ORGANIZATION FOR NUCLEAR RESEARCH (CERN)}
\vspace*{1.5cm}
\noindent
\begin{tabular*}{\linewidth}{lc@{\extracolsep{\fill}}r@{\extracolsep{0pt}}}
\ifthenelse{\boolean{pdflatex}}
{\vspace*{-2.7cm}\mbox{\!\!\!\includegraphics[width=.14\textwidth]{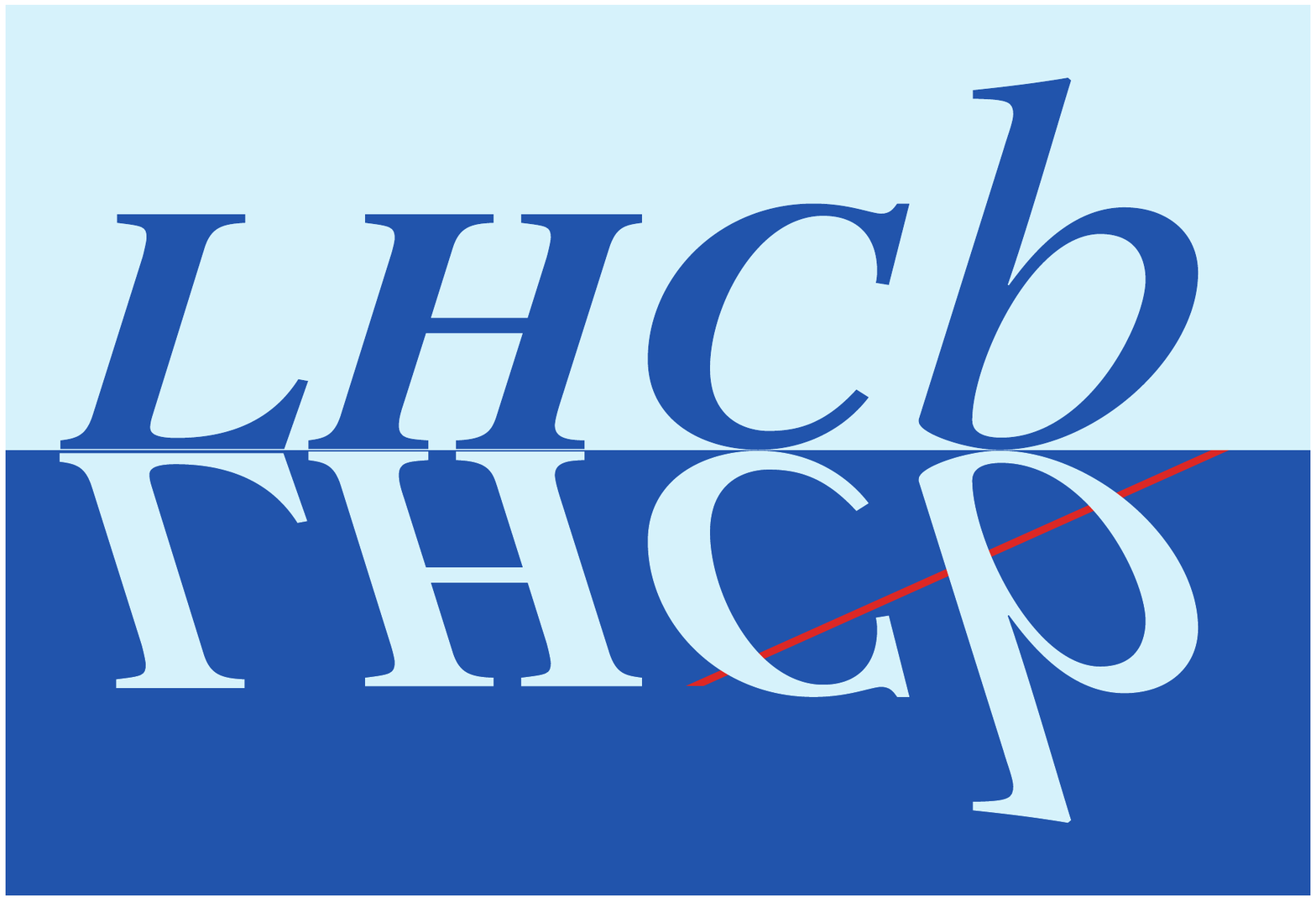}} & &}%
{\vspace*{-1.2cm}\mbox{\!\!\!\includegraphics[width=.12\textwidth]{lhcb-logo.eps}} & &}%
\\
 & & CERN-EP-2016-053 \\  
 & & LHCb-PAPER-2015-050 \\  
 & & April 21, 2016 \\ 
 & & \\
\end{tabular*}

\vspace*{3.0cm}

{\normalfont\bfseries\boldmath\huge
\begin{center}
  Observation of $\Bs\to\Dzb\KS$ and evidence for $\Bs\to\Dstarzb\KS$ decays
\end{center}
}

\vspace*{.5cm}

\begin{center}
The LHCb collaboration\footnote{Authors are listed at the end of this letter.}
\end{center}

\vspace{\fill}

\begin{abstract}
\noindent
The first observation of the $\Bs\to\Dzb\KS$ decay mode and evidence for the $\Bs\to\Dstarzb\KS$ decay mode are reported.
The data sample corresponds to an integrated luminosity of 3.0\invfb collected in $pp$ collisions by \lhcb at center-of-mass energies of 7 and 8\tev.
The branching fractions are measured to be
\begin{align*}
\BR(\Bs\to\Dzb\Kzb) &= (4.3\pm0.5\stat\pm0.3\syst\pm0.3\systfsfd\pm0.6\systext)\times10^{-4},\\ \BR(\Bs\to\Dstarzb\Kzb) &=  (2.8\pm1.0\stat\pm0.3\syst\pm0.2\systfsfd\pm0.4\systext)\times10^{-4},
\end{align*}
 where the uncertainties are due to contributions coming from statistical precision, systematic effects, and the precision of two external inputs, the ratio $f_s/f_d$ and the branching fraction of $\Bz\to\Dzb\KS$, which is used as a calibration channel.

\end{abstract}

\vspace*{1.0cm}

\begin{center}
  Published in Phys.~Rev.~Lett.~116,~161802~(2016)
\end{center}

\vspace{\fill}

{\footnotesize
\centerline{\copyright~CERN on behalf of the \lhcb collaboration, licence \href{http://creativecommons.org/licenses/by/4.0/}{CC-BY-4.0}.}}
\vspace*{2mm}

\end{titlepage}


\newpage
\setcounter{page}{2}
\mbox{~}

\cleardoublepage


\renewcommand{\thefootnote}{\arabic{footnote}}
\setcounter{footnote}{0}



\pagestyle{plain} 
\setcounter{page}{1}
\pagenumbering{arabic}


%



\noindent The study of \CP violation is one of the most important topics in flavor physics.
In \Bd decays, the phenomenon of \CP violation has been extensively studied at BaBar, Belle and \lhcb, which confirmed many predictions of the Standard Model (SM)~\cite{Aubert:2001nu,Abe:2001xe,LHCb-PAPER-2013-020,Bevan:2014iga}.
Nowadays, the focus is on the search for Beyond the Standard Model (BSM) effects by improving the statistical precision of the \CP violation parameters and looking for deviations from the SM predictions.

In the SM, violation of \CP symmetry in \B decays is commonly parameterized by three phase angles ($\alpha$, $\beta$, $\gamma$) derived from the Cabibbo-Kobayashi-Maskawa matrix, which describes the charged-current interactions among quarks~\cite{Kobayashi:1973fv}.
Since the angles sum up to 180\degrees, any deviation found in measurements of the phases would be a sign of BSM physics affecting at least one of the results.
Currently the angle $\gamma$ is only known with an uncertainty of about 10\degrees~\cite{HFAG};
experimental efforts are required to improve its precision and thus the sensitivity to BSM effects.
Another sensitive observable is the \Bs mixing phase, $\phi_s$, which in the SM is predicted with good precision to be close to zero~\cite{Charles:2011va}.
Any significant deviation here would also reveal physics BSM~\cite{Buras:2009if,Chiang:2009ev}.
The current uncertainty is $\mathcal{O}(0.1)\rad$~\cite{HFAG}.

In this Letter, two decay modes that can improve the knowledge of $\gamma$ and $\phi_s$ are studied.
The \decay{\Bd}{\Dzb\KS} decay\ifthenelse{\boolean{prlstyle}}{~}{}\footnote{Unless otherwise specified, the inclusion of charge conjugate reactions is implied.} offers a determination of the angle $\gamma$ with small theoretical uncertainties~\cite{Gronau:2004io}, while \decay{\Bs}{\Dstarzbprime\KS}, similar to the \decay{\Bs}{\Dstarzbprime\phi}~\cite{LHCb-PAPER-2013-035} mode, provides sensitivity to $\phi_s$ with a theoretical accuracy of $\mathcal{O}(0.01)\rad$~\cite{Fleischer:2003ai}.

While the decay \decay{\Bd}{\Dstarzbprime\KS} has been seen at the \B factories~\cite{Aubert:2006qn}, \decay{\Bs}{\Dstarzbprime\KS} decays have not previously been observed.
Theoretical predictions of their branching fractions are of the order of $5\times10^{-4}$\cite{Colangelo:2005hh,Chua:2007qw,Chiang:2007bd}.
This Letter reports the first observation of \decay{\Bs}{\Dzb\KS} and evidence for \decay{\Bs}{\Dstarzb\KS} decays, and provides measurements of branching fractions of these channels normalized to \decay{\Bd}{\Dzb\KS} decays.


The analysis is based on data collected in $pp$ collisions by the \lhcb experiment at $\sqrt{s}=7\tev$ and 8\tev corresponding to an integrated luminosity of 3.0\invfb.
The \lhcb detector~\cite{Alves:2008zz,LHCb-DP-2014-002} is a single-arm forward
spectrometer covering the \mbox{pseudorapidity} range $2<\eta <5$,
designed for the study of particles containing \bquark or \cquark
quarks. The detector includes a high-precision tracking system
consisting of a silicon-strip vertex detector surrounding the $pp$
interaction region, a large-area silicon-strip detector located
upstream of a dipole magnet with a bending power of about
$4{\mathrm{\,Tm}}$, and three stations of silicon-strip detectors and straw
drift tubes placed downstream of the magnet.
The tracking system provides a measurement of momentum, \ptot, of charged particles with
a relative uncertainty that varies from 0.5\% at low momentum to 1.0\% at 200\gevc.
Two ring-imaging Cherenkov (RICH) detectors are able to discriminate between different species of charged hadrons.
The online event selection is performed by a trigger,
which consists of a hardware stage, based on information from the calorimeter and muon
systems, followed by a software stage, which applies a full event
reconstruction.

In the simulation, $pp$ collisions are generated using
\pythia~\cite{Sjostrand:2006za,*Sjostrand:2007gs}
 with a specific \lhcb
configuration~\cite{LHCb-PROC-2010-056}.  Decays of hadronic particles
are described by \evtgen~\cite{Lange:2001uf}, in which final-state
radiation is generated using \photos~\cite{Golonka:2005pn}. The
interaction of the generated particles with the detector, and its response,
are implemented using the \geant
toolkit~\cite{Allison:2006ve, *Agostinelli:2002hh} as described in
Ref.~\cite{LHCb-PROC-2011-006}.


At the hardware trigger stage, events are required to have a muon with high \pt or a
hadron, photon or electron with high transverse energy deposited in the calorimeters.
The software trigger requires a two-, three- or four-track
secondary vertex with a significant displacement from any reconstructed primary vertex (PV).
At least one of these tracks must have $\pt > 1.7\gevc$ and be inconsistent with originating from a PV.
A multivariate algorithm~\cite{BBDT} is used to identify secondary vertices consistent with the decay
of a \bquark hadron.

Candidate \decay{\KS}{\pip\pim} decays are reconstructed in two different categories:
the first involving \KS mesons that decay early enough for the
daughter pions to be reconstructed in the vertex detector, referred to as \emph{long}; and the
second containing \KS that decay later such that track segments of the
pions cannot be formed in the vertex detector, referred to as \emph{downstream}.
The long category has better mass, momentum and vertex resolution than the
downstream category.
Long (downstream) \KS candidates are required to have decay lengths larger than 12 (9) times the decay length uncertainty.
The invariant mass of the candidate is required to be within 30\mevcc of the known \KS mass~\cite{PDG2014}.

The \decay{\Dzb}{\Kp\pim} candidates are formed from combinations of kaon and pion candidate tracks identified by the RICH detectors.
The pion (kaon) must have $p>1~(5)\gevc$ and $\pt>100~(500)\mevc$, and be inconsistent with originating from a PV.
The invariant mass of the candidate is required to be within 50\mevcc of the known \Dzb mass~\cite{PDG2014}.

The \B (\Bz or \Bs) candidate is formed by combining \Dzb and \KS candidates and requiring an invariant mass in the range 4500--7000\mevcc, a decay time greater than 0.2\ps and a momentum vector pointing back to the associated PV.
To improve the mass resolution of the \B candidates, a kinematic fit is performed constraining the masses of the \Dzb and \KS candidates to the known values~\cite{PDG2014}.

The purity of the \B candidate sample is then increased by means of a multivariate classifier~\cite{Breiman,AdaBoost} that separates signal from combinatorial background.
Separate algorithms are trained for candidates with long and downstream \KS candidates.
The discriminating variables used in the classifier are the \chisq of the kinematic fit, geometric variables related to the finite lifetime of the \B, \Dzb and \KS, the decay time, \pt and $p$ of the \KS candidate.
The multivariate classifier is trained and tested using signal candidates from simulations and background candidates from data in the upper sideband of the \B mass spectrum, corresponding to $m(\Dzb\KS)>5500\mevcc$, where no backgrounds are expected from \B decays in which a photon or a $\pi$ meson is not reconstructed.
The selection is optimized to minimize the statistical uncertainty on the ratio of \Bs over \Bd signal event yields.
The signal efficiency and background rejection factors are 76\% and 98\%, respectively.
\B candidates in the mass range 5000--5900\mevcc are retained.
Multiple candidates occur in  0.2\% (0.4\%) of long (downstream) \KS events in which case one candidate, chosen at random, is kept.


The \Bs and \Bd signal yields in the selected sample are obtained from an unbinned extended maximum likelihood fit simultaneously performed on the long and downstream \KS samples.
The observables used in the fit are $m_{\KS}$, the mass of the \decay{\KS}{\pip\pim} candidates, $m_{\Dzb}$, the mass of the \decay{\Dzb}{\Kp\pim} candidates, and $m_{\B}$, the mass of the \B meson candidates.
The probability density function (PDF) contains four terms
\begin{align}
  \mathcal{P}(m_{\Dzb},m_{\KS},m_{\B}) =
  \sum_{i=1}^4 N_i \cdot \mathcal{F}_i(m_{\Dzb},m_{\KS},m_{\B}) =\ifthenelse{\boolean{prlstyle}}{\nonumber\\=}{}
  \sum_{i=1}^4 N_i \cdot \mathcal{P}_i(m_{\B}) \cdot \mathcal{S}_i(m_{\Dzb},m_{\KS})
\end{align}
where $N_i$ represents the respective yield, $\mathcal{P}_i$ parametrizes the mass distribution of the \B meson candidates and $\mathcal{S}_i$ is the joint PDF of the candidates for its decay products.
The term $\mathcal{F}_1$ describes correctly reconstructed \Dzb and \KS candidates, $\mathcal{F}_2$ a correctly reconstructed \Dzb meson in association with two random pions, $\mathcal{F}_3$ a correctly reconstructed \KS meson in association with a random kaon and pion, and $\mathcal{F}_4$ random combinations of the four final state particles.
Johnson SU distributions~\cite{JohnsonSU}, characterized by asymmetric tails to account for radiative losses and vertex reconstruction uncertainties, are used to parametrize the \Dzb and \KS signals in $\mathcal{S}_{1,2,3}$ and exponential functions describe the backgrounds in $\mathcal{S}_{2,3,4}$.

The \B mass in candidates with correctly reconstructed \Dzb and \KS mesons ($\mathcal{P}_1$) is described by three categories of shapes: \decay{\BorBs}{\Dzb\KS} signal, peaking structures at lower mass from other \B decays and combinatorial background.
Signal shapes for the \Bd and \Bs candidates decaying to $\Dzb\KS$ are described by means of Johnson SU distributions with shape parameters determined from fits to the simulated signal samples, corrected for differences between the simulation and data.
The peaking structures at lower mass correspond to decays of \Bd and \Bs mesons that include \Dzb and \KS mesons in the final state where a photon or a $\pi$ meson is not reconstructed, such as \decay{B^0_{(s)}}{\Dstarzb(\Dzb\piz)\KS}, \decay{B^0_{(s)}}{\Dstarzb(\Dzb\gamma)\KS}, \decay{\Bp}{\Dzb\Kstarp(\KS\pip)}, and \decay{B^0_{(s)}}{\Dzb\Kstarz(\KS\piz)}.
These shapes are described with kernel estimated PDFs~\cite{Cranmer:2000du} obtained from simulation.

The same exponential function is used for the combinatorial background description of the \B mass distribution in $\mathcal{P}_{1,2,3,4}$.
Possible contaminations from \decay{B^0_{(s)}}{\Dzb\pip\pim} and \decay{B^0_{(s)}}{\Dstarzb\pip\pim} in $\mathcal{P}_{2}$, and \decay{B^0_{(s)}}{\KS\Kp\pim} and \decay{B^0_{(s)}}{\Kstarz(\KS\piz)\Kp\pim} in $\mathcal{P}_{3}$ are accounted for using the function that describes the $B^{0}_{(s)}$ candidates in $\mathcal{P}_{1}$.

The PDFs $\mathcal{F}_i$ are distinct for the long and downstream samples, but share certain parameters including those of the \Dzb signal distribution and the yield fractions of the non-combinatorial components of the \B mass spectrum.
Gaussian constraints are applied to the branching fraction ratios $\BR(\decay{\Bs}{\Dzb\Kstarz})/[\BR(\decay{\Bz}{\Dzb\Kstarz})+\BR(\decay{\Bs}{\Dzb\Kstarz})]$ and $\BR(\decay{\BorBs}{\Dstarzb(\Dzb\piz)\Kz})/[\BR(\decay{\BorBs}{\Dstarzb(\Dzb\gamma)\Kz})+\BR(\decay{\BorBs}{\Dstarzb(\Dzb\piz)\Kz})]$.
These constraints improve the stability of the fit and are determined from measurements of branching fractions reported in Ref.~\cite{PDG2014}, corrected by the efficiencies of the relevant \BorBs decays as determined from simulated samples.

Projections of the fit results on the data sample are shown in Fig.~\ref{fig:fitresults}.
\ifthenelse{\boolean{prlstyle}}{\begin{figure*}[t]}{\begin{figure}[t]}
\centering
\includegraphics[width=0.24\textwidth]{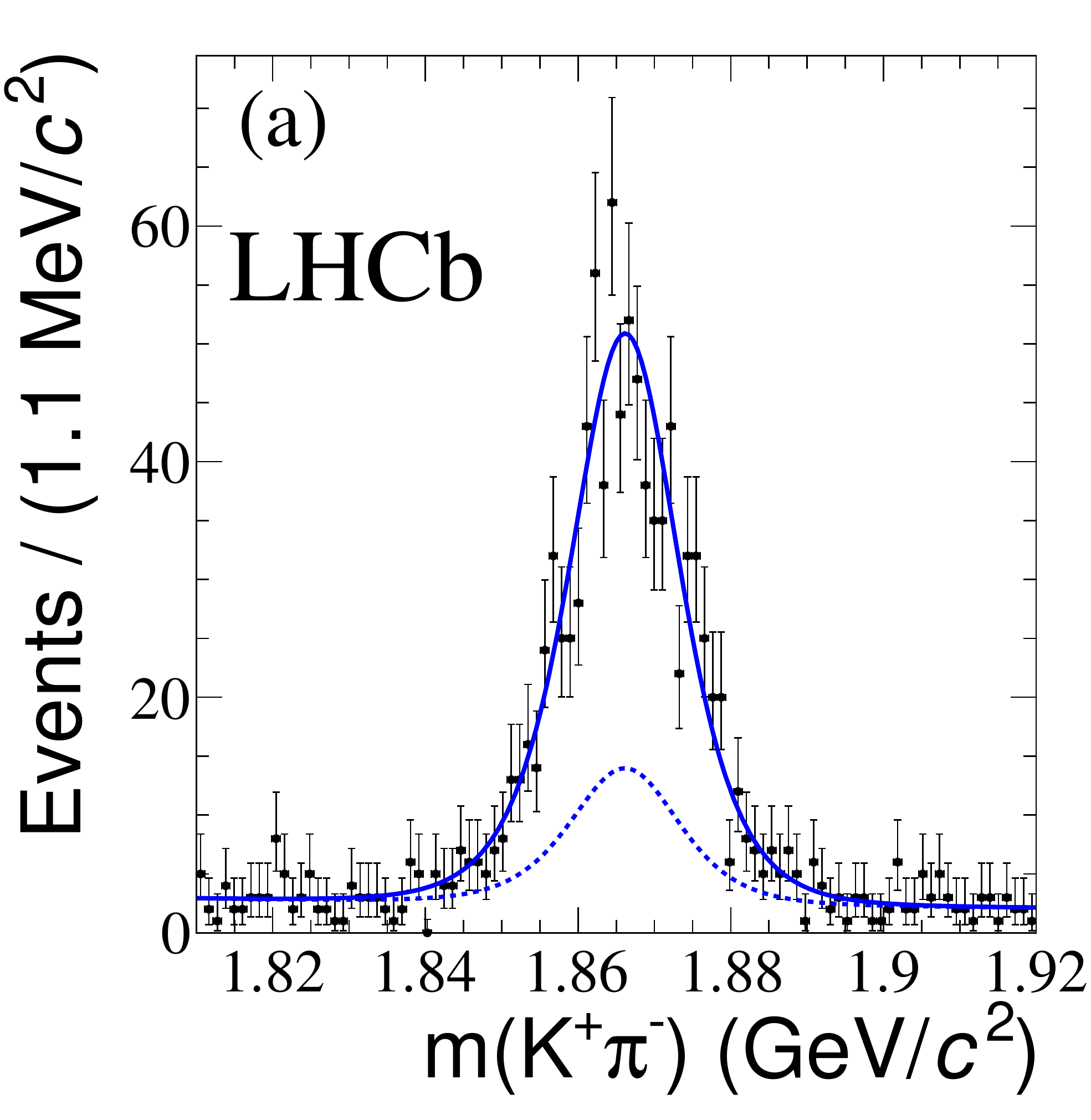}
\includegraphics[width=0.24\textwidth]{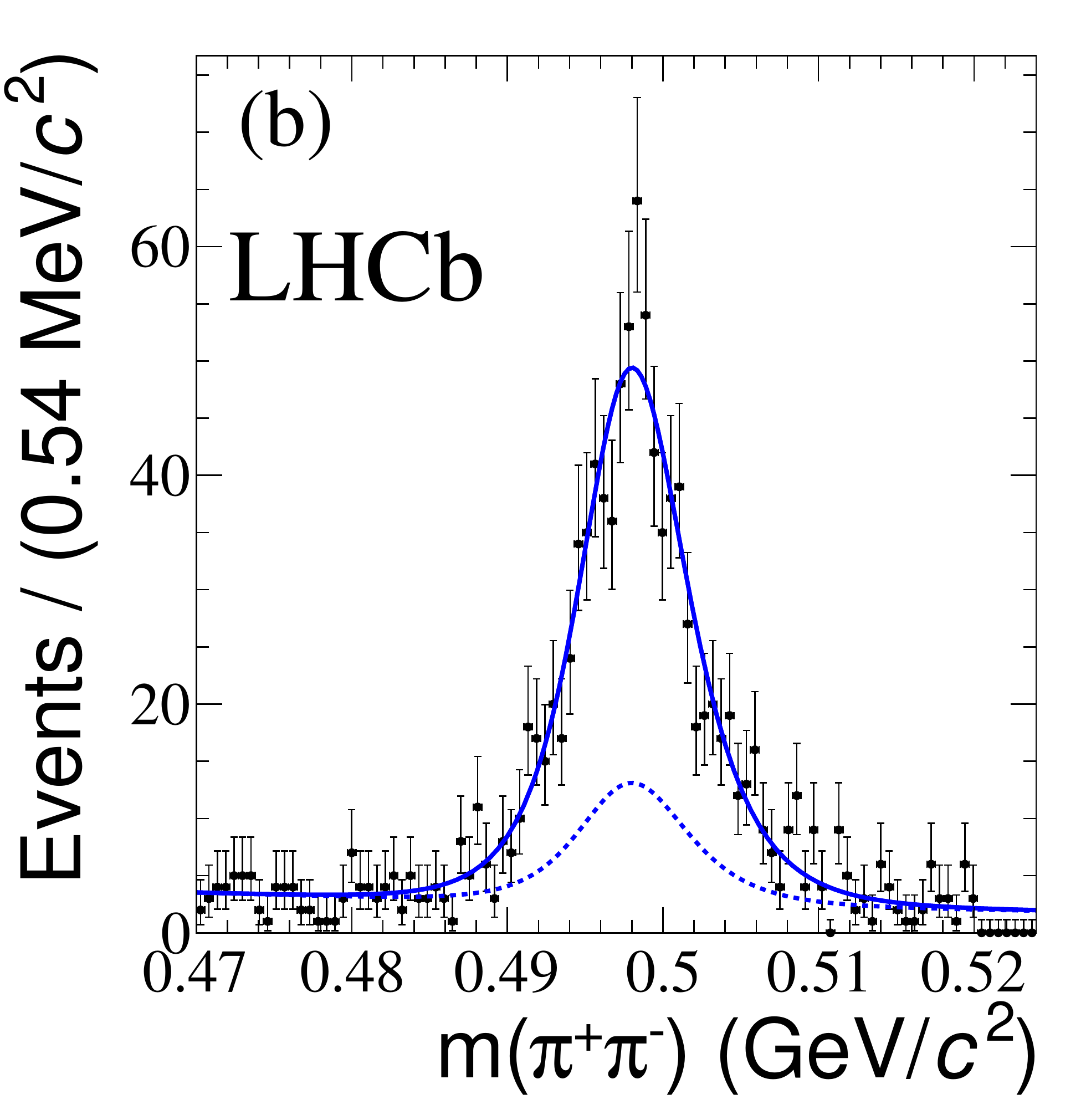}
\includegraphics[width=0.24\textwidth]{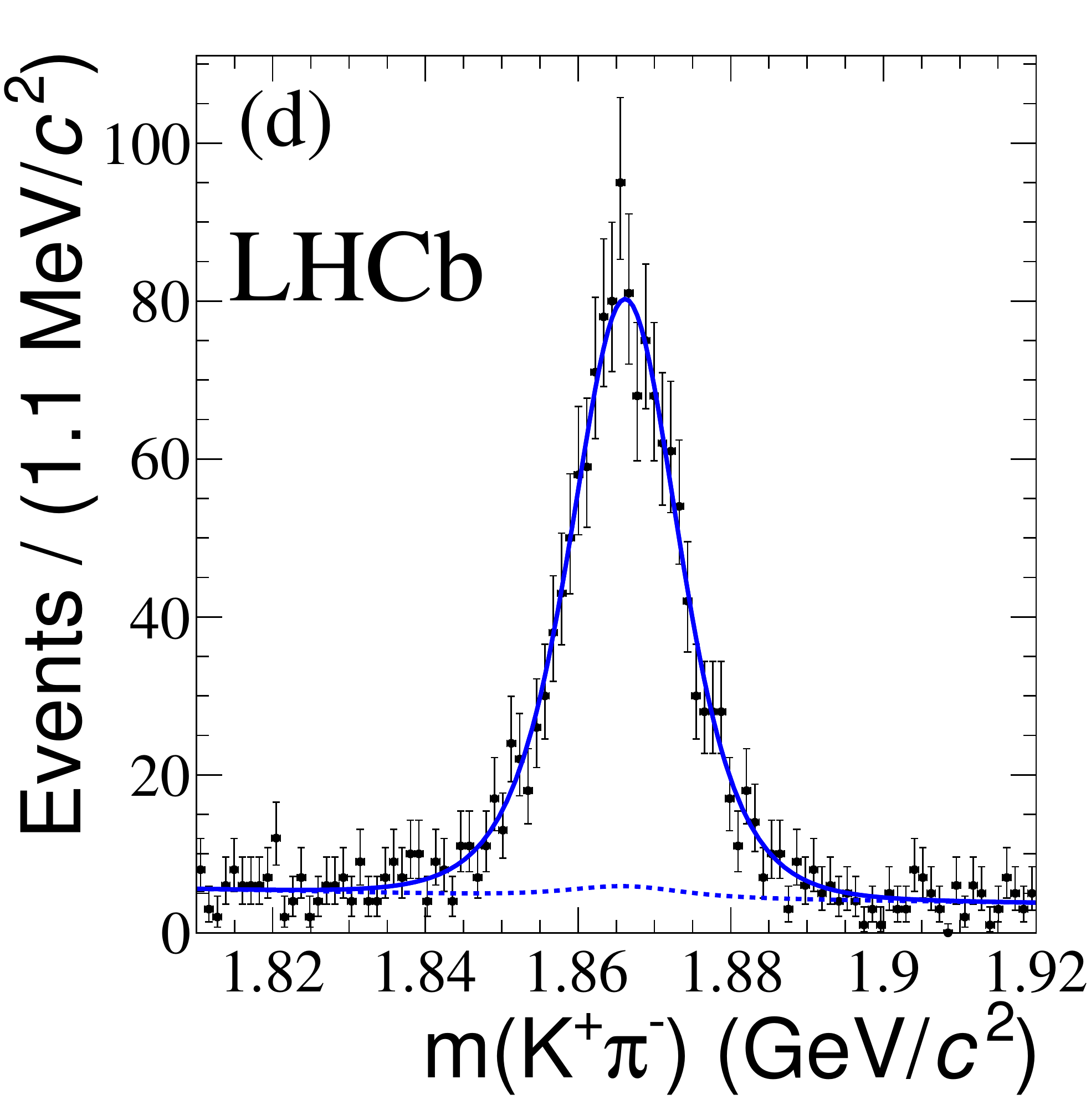}
\includegraphics[width=0.24\textwidth]{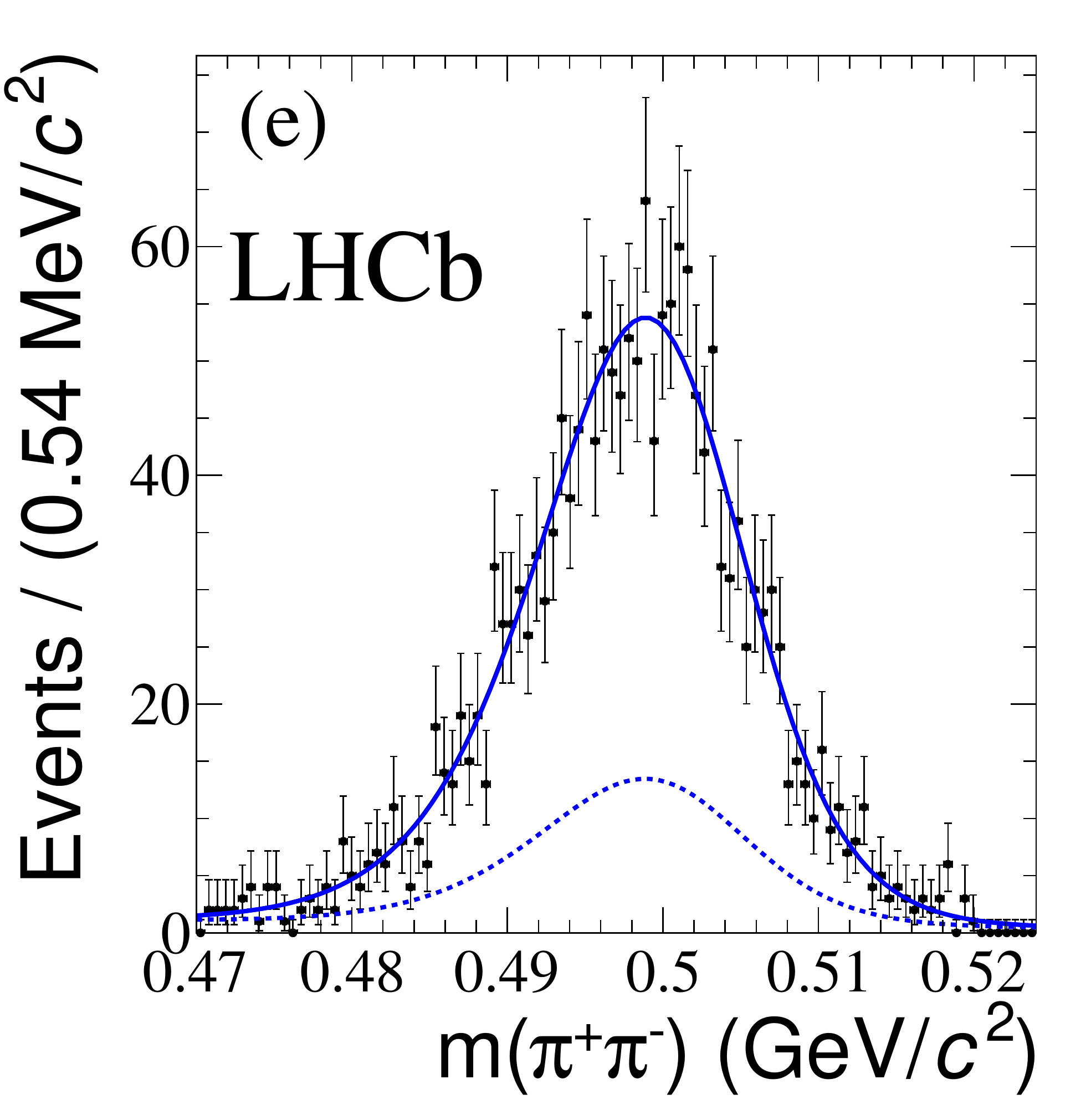}
\includegraphics[width=0.49\textwidth]{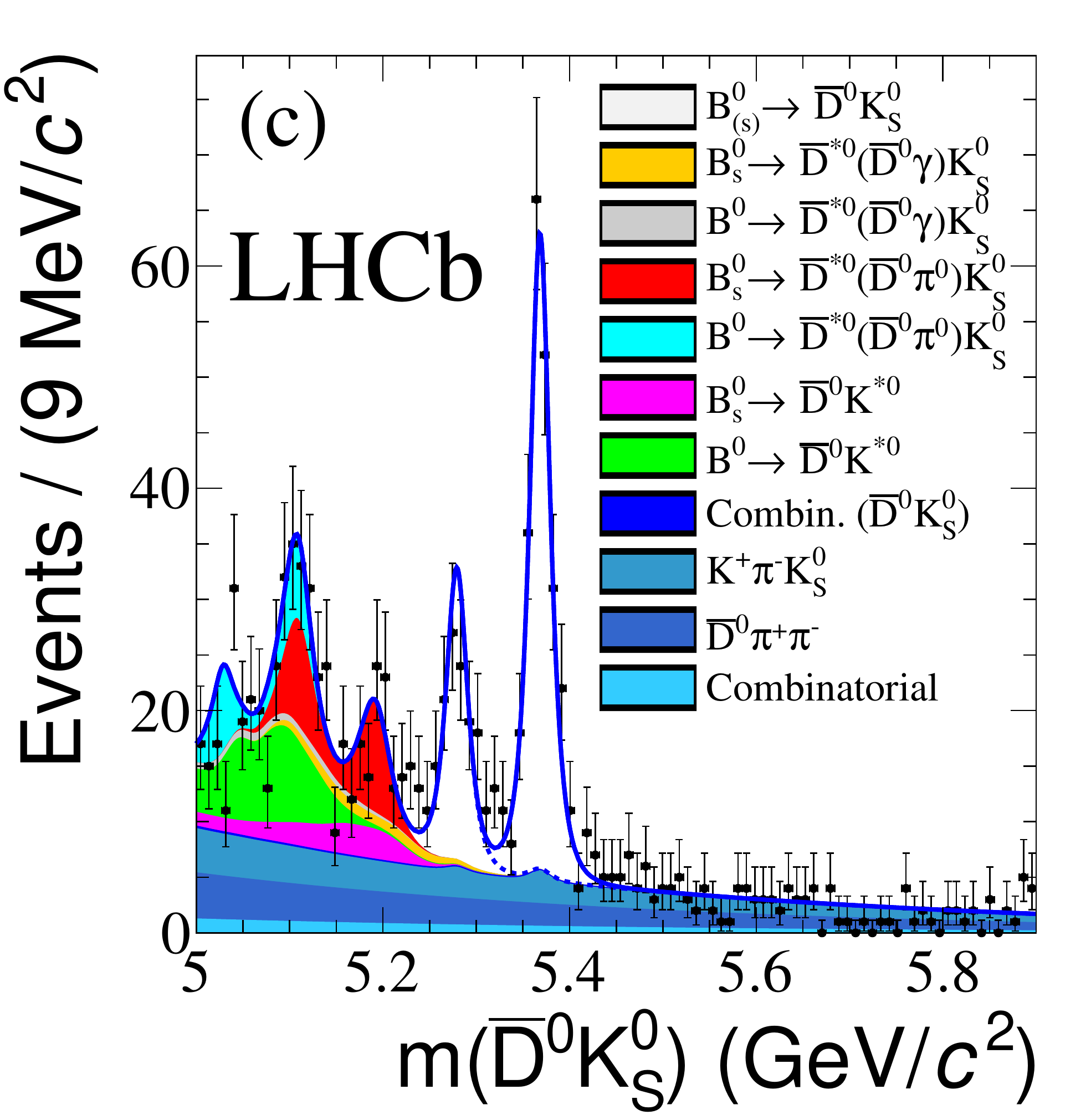}
\includegraphics[width=0.49\textwidth]{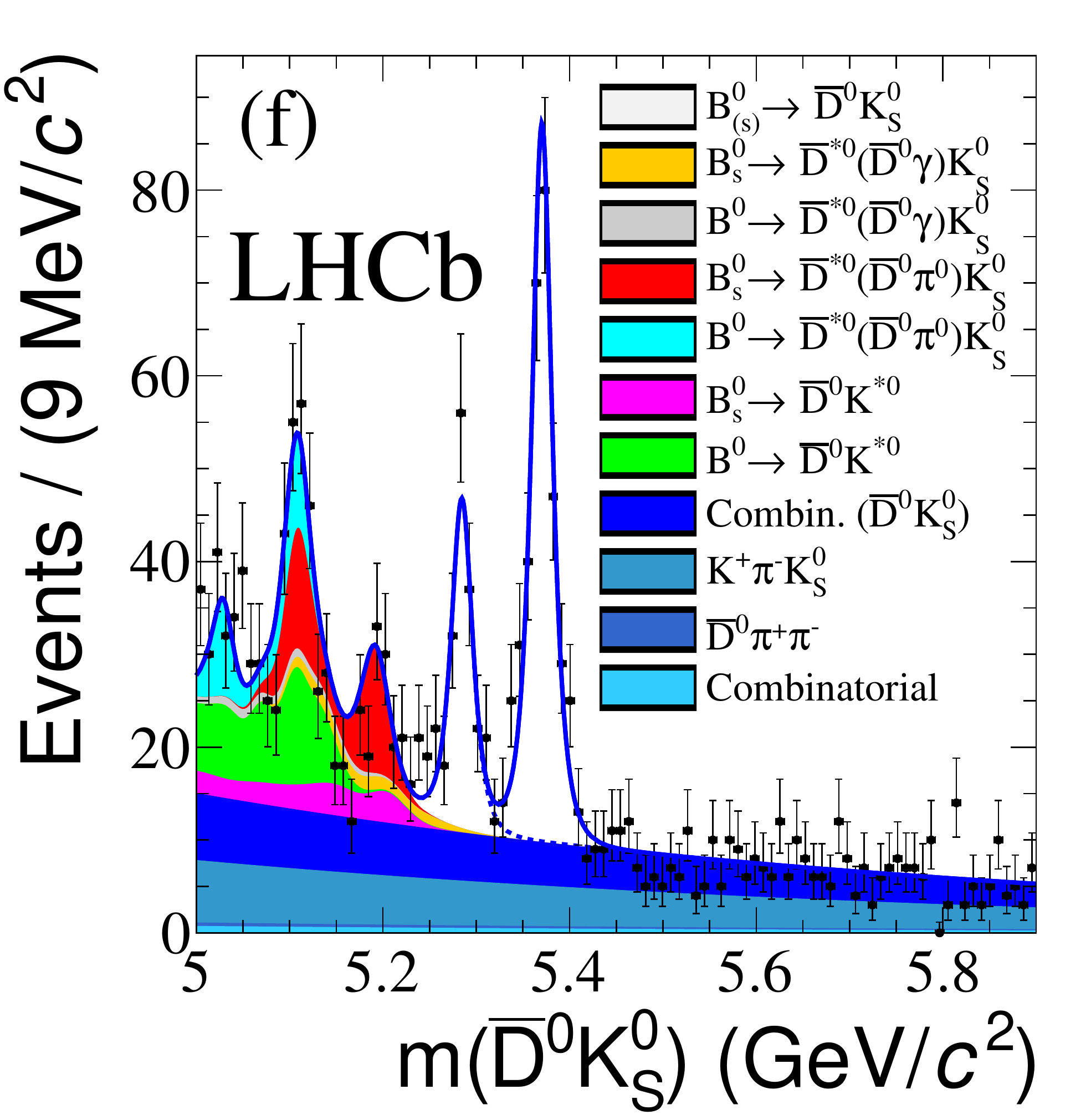}
\caption{\small\label{fig:fitresults}The projection of the fit results (solid line) on the data sample (points) is shown for the \Dzb candidate (a,d), the \KS candidate (b,e) and \B candidate (c,f) mass spectra.
The long \KS sample is shown in (a,b,c), and the downstream sample in (d,e,f).
The dashed line in the \Dzb and \KS candidate mass plots represents events corresponding to background categories $\mathcal{S}_{2,3,4}$ in the fit, and includes peaks due to, for example, real \Dzb mesons paired with two random pions.
The double-peak behavior of the \decay{\BorBs}{\Dstarzb(\Dzb\piz)\KS} shape is due to the missing momentum of the \piz and the helicity amplitude of the \decay{\Dstarzb}{\Dzb\piz} decay.}
\ifthenelse{\boolean{prlstyle}}{\end{figure*}}{\end{figure}}
The numbers of signal candidates determined from the fit are $N(\Bd\to\Dzb\KS) = 219\pm21$, $N(\Bs\to\Dzb\KS) = 471\pm26$ and $N(\Bs\to\Dstarzb\KS) = 258\pm83$, where the uncertainties are purely statistical.

The branching fractions, \BR, of the \decay{\Bs}{\Dstarzbprime\Kzb} decays are calculated from the ratio of branching fractions between \Bs and \Bd
\begin{align}
\BR(\Bs\to\Dstarzbprime\Kzb) &= \mathcal{R}^{(*)} \times \Bsum
\end{align}
where $\Bsum = \BR(\decay{\Bd}{\Dzb\Kz}) + \BR(\decay{\Bdb}{\Dzb\Kzb})$, since the analysis does not distinguish between \Kz and \Kzb.
The quantity
\begin{align}
\mathcal{R}^{(*)}&= \frac{f_d}{f_s} \frac{N(\Bs\to\Dstarzbprime\KS)}{N(\Bz\to\Dzb\KS)+N(\Bzb\to\Dzb\KS)}\frac{\epsilon_{\Bd}}{\epsilon_{\Bs}}
\end{align}
is the product of the production ratio of \Bd over \Bs decays in \lhcb ($f_d/f_s$), the ratio of reconstructed \Bs and \Bd signal candidates, and the ratio of efficiencies of \Bd to \Bs candidates decaying to $\Dstarzbprime\KS$ in the \lhcb detector ($\epsilon_{\Bd}/\epsilon_{\Bs}$).
The value of $f_s/f_d=0.259\pm0.015$ is provided by previous \lhcb measurements~\cite{LHCb-PAPER-2012-037,LHCb-CONF-2013-011}.
The ratios of efficiencies $\epsilon_{\Bd\to\Dzb\KS}/\epsilon_{\Bs\to\Dzb\KS}=0.997\pm0.024$ and $\epsilon_{\Bd\to\Dzb\KS}/\epsilon_{\Bs\to\Dstarzb\KS}=1.181\pm0.029$ are obtained from simulated samples.
The ratio of \Bs and \Bd signal candidates is a free parameter in the fit and is measured to be
$N(\Bs\to\Dzb\KS)/[N(\Bz\to\Dzb\KS)+N(\Bzb\to\Dzb\KS)]=2.15\pm0.23$.
Similarly, the ratio
$N(\Bs\to\Dstarzb\KS)/[N(\Bz\to\Dzb\KS)+N(\Bzb\to\Dzb\KS)]=1.17\pm0.44$
is measured.

Various sources of systematic uncertainty have been considered.
These are summarized in Table~\ref{tab:systematics} and discussed below.
\begin{table}[b]
\centering
\caption{\small\label{tab:systematics}Summary of the systematic uncertainties.}
\begin{tabular}{lcc}
\hline\hline
Source & \decay{\Bs}{\Dzb\KS}& \decay{\Bs}{\Dstarzb\KS}\\
\hline
Fit model & $5.4\%$ & $11.9\%$\\
$\epsilon_{\Bd}/\epsilon_{\Bs}$ & $2.4\%$ & $2.5\%$\\
$f_s/f_d$	& \multicolumn{2}{c}{$5.8\%$}\\
\Bsum & \multicolumn{2}{c}{$13.5\%$}\\
\hline\hline
\end{tabular}
\end{table}%

The uncertainty associated to the fit model is assessed by the use of other functions for the PDFs $\mathcal{P}_i$ and $\mathcal{S}_i$.
For the mass distribution of the signal events, four alternative models are used.
Each pseudoexperiment generated in this way is then fitted with the baseline model and the difference of the signal yields ratio with respect to the generated value is considered.
The mean of the distribution that shows the largest deviation from zero is taken as the systematic uncertainty, corresponding to 5.4\% (11.9\%) for \decay{\Bs}{\Dzb\KS} (\decay{\Bs}{\Dstarzb\KS}).

The ratio of efficiencies of the \Bd and \Bs decays is determined from simulation and is limited by the finite size of the sample.
The statistical uncertainties on the efficiency ratios and the statistical uncertainties of the external inputs, $f_s/f_d$ and the branching fraction \Bsum, are propagated to the systematic uncertainty of this measurement.

To test the stability of the result with respect to the offline selection, the measurement is repeated at different selection cuts on the multivariate classifier.
The deviations from the nominal result are consistent with statistical fluctuations and no systematic uncertainty is assigned.
Possible bias due to the random removal of multiple candidates is tested by removing or keeping all of them, and no significant effect is observed.

Further cross-checks on the stability of the result are made by measuring the branching fractions independently for the long and downstream \KS samples, for the two different polarities of the \lhcb magnet and for different running conditions.
No significant effect is observed.

Only the fit model is considered when determining the systematic uncertainty on the number of signal candidates.
The statistical uncertainty on the efficiencies and on $f_s/f_d$ are also included in the sum in quadrature to give the systematic uncertainty on the ratio of branching fractions $\mathcal{R}^{(*)}$.
Finally, the uncertainty on \Bsum is also included for the measurement of the branching fraction $\BR(\Bs\to\Dstarzbprime\Kzb)$.

Signal yields of
\begin{align*}
  N(\decay{\Bd}{\Dzb\KS}) &= 219 \pm 21\stat \pm 11\syst,\\
  N(\decay{\Bs}{\Dzb\KS}) &= 471 \pm 26\stat \pm 25\syst,\\
  N(\decay{\Bs}{\Dstarzb\KS}) &= 258 \pm 83\stat \pm 30\syst
\end{align*}
are found.
Those results correspond to the first observation of the \decay{\Bs}{\Dzb\KS} decay with a significance of 13.1 standard deviations and evidence for \decay{\Bs}{\Dstarzb\KS} with a significance of 4.4 standard deviations, where the significances are calculated using Wilks' theorem~\cite{Wilks:1938dza}.

The ratios of the branching fractions are
\begin{align*}
\mathcal{R} &= 8.3 \pm 0.9\stat \pm 0.5\syst \pm 0.5\systfsfd,\\
\mathcal{R}^* &= 5.4 \pm 2.0\stat \pm 0.7\syst \pm 0.3\systfsfd.
\end{align*}
Here the correlation coefficient between the two statistical uncertainties is 68\% and that between the two systematic uncertainties is 49\%.
Using the branching fraction $\Bsum=(5.2\pm0.7)\times10^{-5}$~\cite{PDG2014}, the values of the branching fractions are
\ifthenelse{\boolean{prlstyle}}{\begin{widetext}}{}
\begin{align*}
\BR(\decay{\Bs}{\Dzb\Kzb}) &= (4.3 \pm 0.5\stat \pm 0.3\syst \pm 0.3\systfsfd \pm 0.6\systext )\times10^{-4},\\
\BR(\decay{\Bs}{\Dstarzb\Kzb}) &= (2.8 \pm 1.0\stat \pm 0.3\syst \pm 0.2\systfsfd \pm 0.4\systext )\times10^{-4},
\end{align*}
\ifthenelse{\boolean{prlstyle}}{\end{widetext}}{}where the last uncertainty is due to the uncertainty on \Bsum.
These results are consistent with theoretical predictions from Refs.~\cite{Colangelo:2005hh,Chua:2007qw,Chiang:2007bd}, when corrections for the difference in width between the \Bs mass eigenstates~\cite{DeBruyn:2012wj} are taken into account.

This Letter reports the first observation of \decay{\Bs}{\Dzb\KS} and first evidence of \decay{\Bs}{\Dstarzb\KS}.
Since the theoretical predictions for these modes have a small uncertainty, future studies with increased statistics and additional \Dzb decay modes will give significant improvements in the determination of $\phi_s$ and $\gamma$.

\section*{Acknowledgements}

\noindent We express our gratitude to our colleagues in the CERN
accelerator departments for the excellent performance of the LHC. We
thank the technical and administrative staff at the LHCb
institutes. We acknowledge support from CERN and from the national
agencies: CAPES, CNPq, FAPERJ and FINEP (Brazil); NSFC (China);
CNRS/IN2P3 (France); BMBF, DFG and MPG (Germany); INFN (Italy);
FOM and NWO (The Netherlands); MNiSW and NCN (Poland); MEN/IFA (Romania);
MinES and FANO (Russia); MinECo (Spain); SNSF and SER (Switzerland);
NASU (Ukraine); STFC (United Kingdom); NSF (USA).
We acknowledge the computing resources that are provided by CERN, IN2P3 (France), KIT and DESY (Germany), INFN (Italy), SURF (The Netherlands), PIC (Spain), GridPP (United Kingdom), RRCKI and Yandex LLC (Russia), CSCS (Switzerland), IFIN-HH (Romania), CBPF (Brazil), PL-GRID (Poland) and OSC (USA). We are indebted to the communities behind the multiple open
source software packages on which we depend.
Individual groups or members have received support from AvH Foundation (Germany),
EPLANET, Marie Sk\l{}odowska-Curie Actions and ERC (European Union),
Conseil G\'{e}n\'{e}ral de Haute-Savoie, Labex ENIGMASS and OCEVU,
R\'{e}gion Auvergne (France), RFBR and Yandex LLC (Russia), GVA, XuntaGal and GENCAT (Spain), Herchel Smith Fund, The Royal Society, Royal Commission for the Exhibition of 1851 and the Leverhulme Trust (United Kingdom).

\addcontentsline{toc}{section}{References}
\setboolean{inbibliography}{true}
\bibliographystyle{LHCb}
\bibliography{main,LHCb-PAPER,LHCb-CONF,LHCb-DP,LHCb-TDR}

\newpage


\centerline{\large\bf LHCb collaboration}
\begin{flushleft}
\small
R.~Aaij$^{39}$, 
C.~Abell\'{a}n~Beteta$^{41}$, 
B.~Adeva$^{38}$, 
M.~Adinolfi$^{47}$, 
A.~Affolder$^{53}$, 
Z.~Ajaltouni$^{5}$, 
S.~Akar$^{6}$, 
J.~Albrecht$^{10}$, 
F.~Alessio$^{39}$, 
M.~Alexander$^{52}$
% ...

{\footnotesize \it
% ...
$ ^{5}$Clermont Universit\'{e}, Universit\'{e} Blaise Pascal, CNRS/IN2P3, LPC, Clermont-Ferrand, France\\
$ ^{6}$CPPM, Aix-Marseille Universit\'{e}, CNRS/IN2P3, Marseille, France\\
% ...
$ ^{10}$Fakult\"{a}t Physik, Technische Universit\"{a}t Dortmund, Dortmund, Germany\\
% ...
$ ^{38}$Universidad de Santiago de Compostela, Santiago de Compostela, Spain\\
$ ^{39}$European Organization for Nuclear Research (CERN), Geneva, Switzerland\\
% ...
$ ^{41}$Physik-Institut, Universit\"{a}t Z\"{u}rich, Z\"{u}rich, Switzerland\\
% ...
$ ^{47}$H.H. Wills Physics Laboratory, University of Bristol, Bristol, United Kingdom\\
% ...
$ ^{52}$School of Physics and Astronomy, University of Glasgow, Glasgow, United Kingdom\\
$ ^{53}$Oliver Lodge Laboratory, University of Liverpool, Liverpool, United Kingdom\\
% ...
}
\end{flushleft}

\end{document}